\documentclass[aps,prl,twocolumn,superscriptaddress,showpacs,floatfix,10pt]{revtex4-1}
\usepackage{amsmath}
\usepackage{graphicx}
\usepackage{color}
\usepackage{epstopdf}
\usepackage{natbib}
\usepackage{xfrac}
\usepackage{wasysym}
\usepackage{multirow}

\newcommand{\kk}{\mathbf{k}}
\newcommand{\rr}{\mathbf{r}}

\newcommand{\EE}{\mathbf{E}}
\newcommand{\BB}{\mathbf{B}}
\newcommand{\En}{\mathcal{E}}

\newcommand{\eref}[1]{Eq.~(\ref{#1})}
\newcommand{\fref}[1]{Fig.~\ref{#1}}

\usepackage{xparse}
\ExplSyntaxOn
\NewDocumentCommand{\mref}{m}{\quinn_mref:n {#1}}
\seq_new:N \l_quinn_mref_seq
\cs_new:Npn \quinn_mref:n #1
 {
  \seq_set_split:Nnn \l_quinn_mref_seq { , } { #1 }
  \seq_pop_right:NN \l_quinn_mref_seq \l_tmpa_tl
  \seq_map_inline:Nn \l_quinn_mref_seq
    { \ref{##1},\nobreakspace } 
  \exp_args:NV \ref \l_tmpa_tl 
 }
\ExplSyntaxOff

\newcommand{\out}[1]{}

\begin{document}
\title{Unified picture for the colossal thermopower compound FeSb$_2$}

\author{M. Battiato}
\affiliation{Institute for Solid State Physics, Vienna University of Technology,  Vienna, Austria}
\email[]{marco.battiato@ifp.tuwien.ac.at}
\author{J.~M. Tomczak}
\affiliation{Institute for Solid State Physics, Vienna University of Technology,  Vienna, Austria}
\author{Z. Zhong}
\affiliation{Institute for Solid State Physics, Vienna University of Technology,  Vienna, Austria}
\author{K. Held}
\affiliation{Institute for Solid State Physics, Vienna University of Technology,  Vienna, Austria}

\date{\today}

\begin{abstract}
We identify the driving mechanism of the gigantic Seebeck coefficient in FeSb$_2$ as the phonon-drag effect 
associated with an in-gap density of states that we demonstrate to derive from excess iron.
We accurately model electronic and thermoelectric transport coefficients and explain the
so far ill-understood correlation of maxima and inflection points in different response functions.
Our scenario has far-reaching consequences for attempts to harvest the spectacular powerfactor of FeSb$_2$.
\end{abstract}


\maketitle

\paragraph{Introduction.}
Thermoelectrics\cite{Zlatic} hold great promise for sustainable energy solutions\cite{Snyder2008} and technological applications such as Peltier coolers, heat pumps\cite{Dresselhaus2013}, 
microscopic generators\cite{PhysRevB.89.125103} and probes for quality control of solid state devices\cite{thermoelectric_imaging}. In recent years, more and more intermetallic compounds with narrow gaps have been found to exhibit large thermoelectrical effects. Among them, the iron antimonide FeSb$_2$\cite{PhysRevB.67.155205,PhysRevB.72.045103,0295-5075-80-1-17008,sun_dalton} is the most prominent. 
In fact, it boasts the largest thermoelectric powerfactor ever measured\cite{0295-5075-80-1-17008,sun_dalton,APEX.2.091102,PhysRevB.86.115121}. The thermopower assumes its maximum at around 15~K, making FeSb$_2$ the prime candidate for thermoelectric cooling devices at cryogenic temperatures. 
The origin of the large thermoelectrical response is, however, unknown to date. While electronic correlation effects have been advocated as main benefactor\cite{0295-5075-80-1-17008,APEX.2.091102,sun_dalton,sun:153308,PhysRevB.72.045103,PhysRevB.86.115121,cava2013fesb2,PhysRevB.88.245203}, other works suggested the so-called phonon-drag effect to cause the colossal thermopower\cite{jmt_fesb2,MRC:8871060}.

Besides the large magnitude of thermoelectricity,  several transport quantities of FeSb$_2$ exhibit a pronounced temperature dependence\cite{0295-5075-80-1-17008,APEX.2.091102,sun_dalton,sun:153308,PhysRevB.86.115121,cava2013fesb2,PhysRevB.88.245203}. In particular there is a correlation of features in different response functions\cite{sun:153308,PhysRevB.88.245203,0953-8984-21-11-113101}: Extrema in the Nernst coefficient correspond to inflection points in both the Seebeck coefficient and the resistivity. 
Likewise, at these characteristic temperatures features appear in the magnetoresistance and the Hall coefficient. 
This intimate linkage between electrical and thermoelectrical quantities heralds a connection to the mechanism behind the colossal Seebeck effect.
Yet, in all, so far no convincing comprehensive physical picture for this distinctive behaviour has been proposed. 

Here, we unveil the origin behind the characteristic temperature scales, as well as the mechanism for the colossal thermoelectricity in FeSb$_2$.
We report a minimal model consisting of
an effective one-particle electronic structure with two crucial ingredients: (1) narrow in-gap states and (2) an electron-phonon coupling mechanism.
Our approach captures all transport phenomena on a quantitative level and
pinpoints a new scenario for thermoelectricity in FeSb$_2$, from which the {\it phonon-drag associated with an in-gap density of states} emerges as the main driver. 
Finally, using first principles calculations, we identify 
a diminutive (intrinsic) off-stoichiometry, Fe$_{1+x}$Sb$_{2-x}$, as likely origin of these in-gap states, in line with the large sample-dependence of experimental observables possibly due to different crystal growth techniques\cite{PhysRevB.67.155205,PhysRevB.72.045103,0295-5075-80-1-17008,APEX.2.091102,sun_dalton,sun:153308,PhysRevB.72.045103,PhysRevB.86.115121,cava2013fesb2,PhysRevB.88.245203}.

\paragraph{Theoretical framework.}
Employing Boltzmann's theory in the relaxation time approximation\cite{Suppl}, we study a variety of electrical and thermoelectrical observables.
For the latter, we explicitly include contributions from the {\it phonon-drag effect}:
A thermal gradient $\nabla_{\rr} T $ produces an out-of-equilibrium population in the phonon Brillouin zone. In case of acoustic phonons, at low temperature the effect is an increase in the population for wave vectors 
opposite the direction of the thermal gradient. This will induce a heat flux. However, scatterings will drive this phonon population back towards thermal equilibrium. 
One mechanism responsible for the latter is the scattering with electrons, a process in which linear momentum is transferred to the electronic subsystem: the phonon-drag effect.
The correct, yet cumbersome, way to treat these scattering events is to add a term to the collision integral in the Boltzmann equation.
Instead, we here propose, as an elegant short-cut, to treat the discrete scatterings as a continuous injection of linear momentum.
It is easy to show that for isotropic acoustic phonon branches, the transferred linear momentum is proportional to $\nabla_{\rr} T$.
Thus an effective description of the phonon drag is achieved by adding a term $\propto \nabla_{\rr} T$ to the semi-classical equation of motion of the electrons
\begin{equation} 
	\dot{\kk}=-\frac{e}{\hbar} \left[ \EE + \dot{\rr} \times \BB  \right] - \alpha \nabla_{\rr} T  \label{eq:equation_motion2}
\end{equation}
which states that the time derivative of the electron wave vector $\kk$ is due to the sum of the external electrical $\EE$ and magnetic  $\BB$ fields, and the contribution from the phonon drag ($\hbar$ is the reduced Planck constant, $e$ the electron's charge and $\rr$ its position at time $t$ satisfying $\dot{\rr}=\frac{1}{\hbar} \nabla_{\kk} \En_i\left( \kk \right)$, where $\En_i$ is the energy dispersion of the $i$-th band). 
For simplicity, we shall use a band and momentum averaged phonon-coupling $\alpha$. 
\begin{figure*}[!t]
 \includegraphics[width=0.99\textwidth]{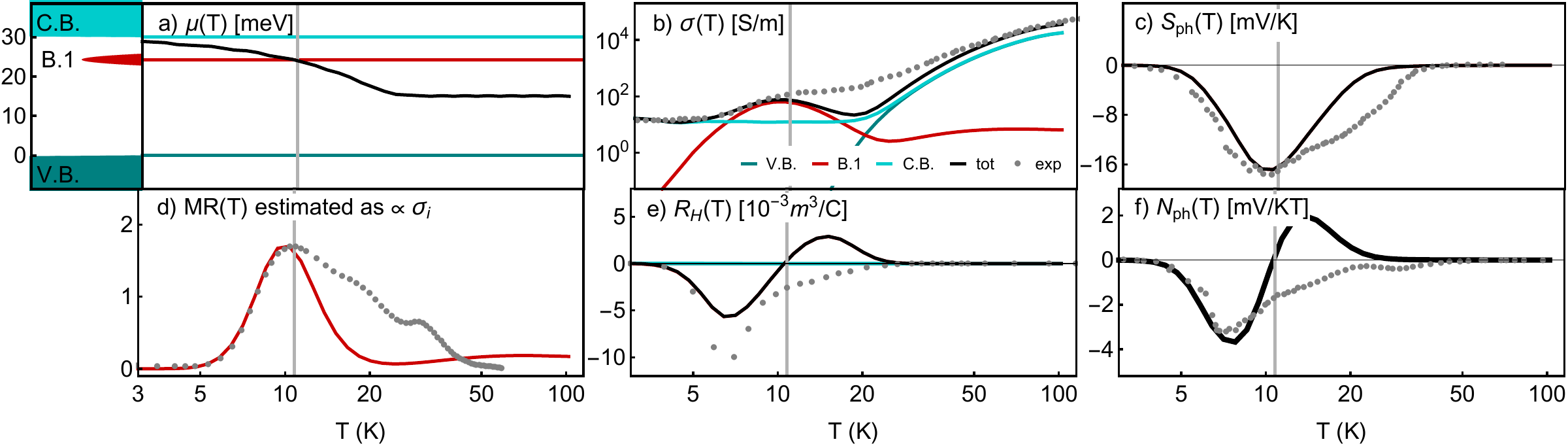}
 \caption{[Colour online] 
Modelled transport properties of FeSb$_2$ with one in-gap state compared to experiment.\cite{PhysRevB.88.245203}
(a) density of states (left) and chemical potential $\mu$ as a function of temperature (right).
(b) electrical conductivity $\sigma$ on a logarithmic scale, consisting of valence (V.B.), conduction (C.B.)
and in-gap band (B.1) contributions.
(c) Seebeck coefficient, (d) magnetoresistance (MR), (e)  Hall coefficient $R_H$, and (f) Nernst coefficient. 
The (gray) vertical line indicates the temperature at which the chemical potential passes the in-gap state $E_1$.
}
 \label{fig:onelevel}
\end{figure*}
Then, the expressions for the phonon-drag driven Seebeck and Nernst coefficient can be readily computed. For instance 
\begin{eqnarray} 
\label{eq:Seebeck}
	S_{ph}  &=&  \frac{ e \alpha }{(2\pi)^3 \, \hbar\, \sigma_{zz}} \sum_{i} 
N_i \tau_i  \int \frac{\partial f}{\partial \En} \left(\frac{\partial \En_i} {\partial k_z}\right)^2 d^3\kk\label{Sph}
\end{eqnarray}
where $f$ is the Fermi function, $\tau_i$ the relaxation time, $N_i$ the band degeneracy, and $\sigma$ the total conductivity.

Evaluating the transport functions requires the knowledge of the dispersions $\En_i$.  
We choose parabolic conduction and valence bands with equal mass.
However, the latter are not enough for a complete description. 
Comparing the experimental resistivity of FeSb$_2$ to the activated behaviour of a perfect semiconductor (see Supplemental Material), we notice that the shoulder around 10K is a clear fingerprint of defect states inside the gap (as also reported in Refs.~\onlinecite{0295-5075-80-1-17008,APEX.2.091102,Takahashi2013}). 
We therefore explicitly include in-gap states in our treatment and consider, as a practical approximation, pairs of spherically symmetric dispersions:  $\En_{i}^\pm\left( k \right) = E_i \pm W_i/2 \cos \left(\sfrac{\pi k}{2\tilde{k}}\right)$, where the reciprocal lattice vector $k$ lies in a Brillouin zone of radius $\tilde{k}$ and $E_i$ is the energy position and $W_i$ the width of the band. 

In the following, it will be important to distinguish between the {\it temperature dependence} and the {\it amplitude} of the response functions. It can be proven\cite{Suppl} that each transport properties can be expressed by an amplitude-factor and a function $\eta$ that contains all non-trivial temperature dependence. For example, the contribution to the phonon-drag thermopower of an in-gap state can be written as: 
 \begin{equation} \label{slow}
	S_{ph}  =   \frac{ \alpha \, W_i^2 \,\tilde{k}_i\, e \; \tau_i\, N_i}{ 192 \hbar^2\, \sigma\, k_B T}  \,\; \eta_{S_{ph}}\!	\!\left(\frac{E_i-\mu(T)}{k_B T} , \frac{W_i}{2k_B T} \right).
\end{equation}
Similar expressions hold for all transport properties, with different amplitude prefactors (depending on $W_i$, $N_i$, $\tau_i$, $\tilde{k}_i$) and characteristic \textit{$\eta$-functions} that all depend
on the same two variables: $\frac{E_i-\mu(T)}{k_B T}$ and $\frac{W_i}{2k_B T}$. The $\eta$-functions can be classified into two categories according to their symmetry with respect to their first argument.
For example, the conductivity, which does not distinguish between electrons and holes will have the same amplitude whether the chemical potential is above or below the band centre, hence
$\eta_{\sigma}$ is {\it even}. 
 Conversely, the Hall coefficient has to change sign  when the chemical potential is at the band centre; $\eta_{R_H}$ is {\it odd} in its first argument. The category of an $\eta$-functions thus reflects the sensitivity of the transport property to the type of carriers. 
A full description can be found in the Supplemental Material\cite{Suppl}, while a quick reference is reported in Table.~\ref{tab:f_functions}. 
It is remarkable that while the electronic Seebeck coefficient is sensitive to the carrier type, the phonon-drag driven one is not. As a consequence the relative symmetry between S, N, R$_H$ and $\sigma$ can be used to distinguish purely electronic from phonon-drag driven thermoelectricity. Interestingly, we also find that all $\eta$-functions are only slowly varying with their second argument. In particular the qualitative shape and the symmetry are inert, and even the peak widths are only weakly dependent on $\frac{W_i}{2k_B T}$. Consequently the {\it temperature profile} of all the transport properties deriving from in-gap states is  {\it entirely} determined by the position $E_i$ of the band with respect to the chemical potential and the absolute temperature. All other parameters merely rescale the {\it amplitude} in a $T$-independent fashion.

\begin{table}
  \begin{tabular}{ | r || c | c | c | c | c |}
  \hline
 $\eta$-function& $\sigma$ & R$_H$ & S & N & MR \\
  \hline \hline
  electronic &\includegraphics[width=0.03\textwidth]{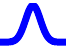} &  \includegraphics[width=0.03\textwidth]{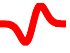}  &  \includegraphics[width=0.03\textwidth]{odd}  &  \includegraphics[width=0.03\textwidth]{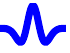} & \includegraphics[width=0.03\textwidth]{even} \\ \hline
  phonon-drag & - & - &  \includegraphics[width=0.03\textwidth]{even} &  \includegraphics[width=0.03\textwidth]{odd} & - \\ \hline
  \end{tabular}
   \caption{Schematic representation of the even/odd character of the $\eta$-functions for {\it in-gap state} derived transport. 
	} 
    \label{tab:f_functions}
\end{table}

\begin{figure*}[!t]
 \includegraphics[width=0.99\textwidth]{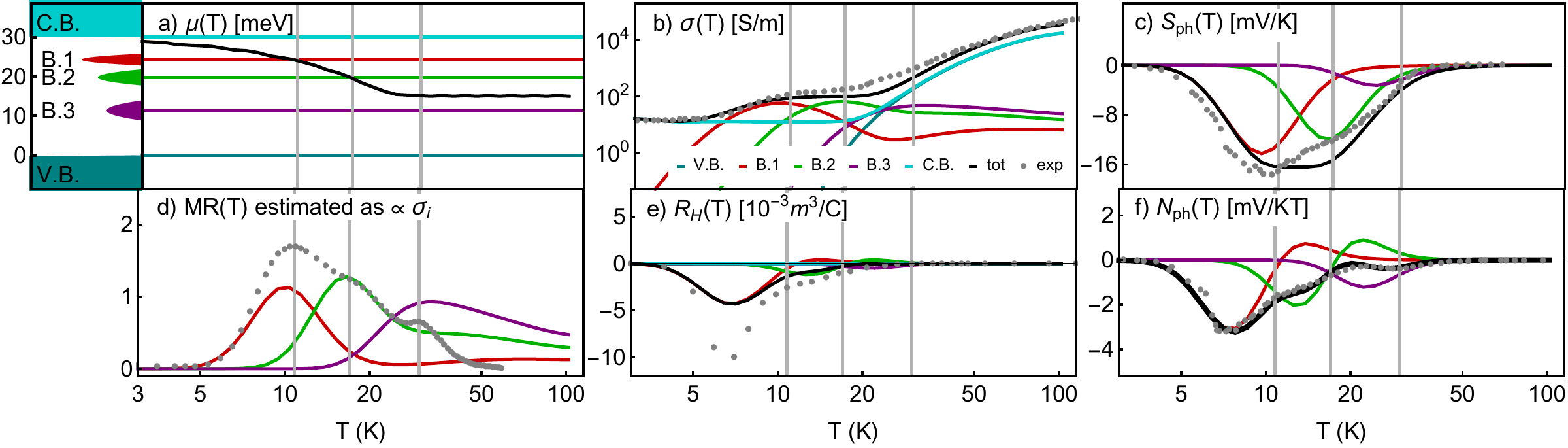}
 \caption{[Colour online] 
Same as Fig.~\ref{fig:onelevel} but for three in-gap states:  $E_i=11.4$, $19.8$, $24.3$ meV.
The (gray) vertical lines are guides to the eye indicating maxima in the MR. For details of the fitting procedure see the text and the Supplemental Material\cite{Suppl}.
}
 \label{fig:threelevels}
\end{figure*}

\paragraph{Temperature dependence of transport properties.}
We now allow for one state $\En_1^\pm(k)$ inside the charge gap, $\Delta=30$meV, to account for the shoulder in the conductivity  (we find that $E_1=24.3$meV yields good agreement, see \fref{fig:onelevel}(a)). All other parameters ($W_i$, $N_i$, $\cdots$) merely describe the relative amplitudes for the conduction/valence and in-gap state contributions, and will be discussed later. 
Here, we first focus on the {\it temperature dependence} of the thermoelectric response to elucidate the ill-understood correlations between features in different transport quantities of FeSb$_2$.
We consider three scenarios:
(1) a purely electronic description of thermoelectricity 
(with contributions from the VB, CB and in-gap states), (2) a dominant phonon-drag effect on valence and conduction carriers, and (3) a preponderant phonon-drag effect that 
couples to electronic in-gap excitations.
Scenario (1) is known to be insufficient to account for the amplitude of thermoelectricity in FeSb$_2$\cite{jmt_fesb2}, but it also yields 
the wrong temperature profile. Scenarios (1) and (2) are dominated by valence and conduction contributions so that the  Seebeck and the Nernst coefficient
have a very comparable temperature dependence and peak positions --at variance with experiments. 
For details of the hence discarded scenarios see the Supplemental Material\cite{Suppl}.

This leaves us with scenario (3), a preponderant phonon-drag effect linked to in-gap states, that we advocate in the following.
In this picture, the shapes of all transport functions are uniquely determined by a single parameter: the position $E_1$ of the in-gap state (see the discussion of \eref{slow}).
As can be seen in \fref{fig:onelevel}, this setup correctly predicts the characteristic peak temperature, as well as the line shapes of the major feature in all five transport functions.

The good agreement notwithstanding, there are deviations above 15K.
To understand this issue, we scrutinize the experimental magnetoresistance (MR). 
We notice that FeSb$_2$ displays a strong response to a magnetic field only in the temperature range where our in-gap band gives a contribution to the conductivity, see \fref{fig:onelevel}.
We do not provide an expression for the unconventional linear-in-B MR of this material, as the underlying mechanism is unresolved. Nonetheless, we can argue that the MR  cannot distinguish between electron and hole transport and has an even-type $\eta$-function. Consequently, the MR peaks at the same temperature as the conductivity (the peak width might be different). In Fig.~\ref{fig:onelevel}d we therefore 
compare the MR to the in-gap-state conductivity and immediately notice that one in-gap band gives only one peak in the MR. 
 However, in the experimental data there are in total three features. 
In Fig.~\ref{fig:threelevels}, we therefore refine our model to include three in-gap states (note that a single state with  an  energy differentiation of its density of states or group velocities can play the same role). All peaks and their correlation in different transport functions are then captured with only three parameters: the energy positions $E_i$ of the in-gap states. To account for relative {\it amplitudes} (see below), our model includes two further parameters that we extract from the ratios of the Nernst coefficient peak amplitudes. 
Then, all five transport coefficients follow accurately: In Fig.~\ref{fig:threelevels} the peak positions and line widths of all features
are  captured\footnote{
The remaining small quantitative discrepancies can be shown to mainly derive from modelling the highly anisotropic FeSb$_2$\cite{0295-5075-80-1-17008,sun_dalton,APEX.2.091102,PhysRevB.86.115121} as isotropic.
}%
. This pinpoints the preponderant influence of in-gap states and their role in the phonon-drag picture as the main driver for the physics of FeSb$_2$.

\paragraph{Amplitude of transport coefficients.}

While we focused so far on the temperature profile of transport coefficients, the modelling in \fref{fig:threelevels} also neatly accounts for the individual amplitudes.
We now detail the parameters that quantitatively determine the response functions. 
The phonon-drag constant can be extracted completely from experiments as $\alpha \approx3 \cdot 10^{13}s^{-1}K^{-1}=229.2\, k_B/\hbar$ at $T=10$K\cite{Suppl}.

Since the temperature dependence of the transport functions is uniquely controlled by the positions of the in-gap states, every transport coefficient only contributes one independent value for the determination of the band-parameters, $W_i$, $N_i$, $\tau_i$. We are thus dealing with a largely under-determined set of equations.
Therefore, we fix some of the parameters of our model to physically reasonable values and then check all remaining adjustable parameters for consistency:
We know that $N_i$ has to be of the order of the unity, so we assign $N_i=1$ for all the in-gap bands.
Then, the width of the Brillouin zone for in-gap states (we assume $\tilde{k}_{1,2,3}=\tilde{k}$) can be extracted from the ratio of the peak amplitudes of the Nernst and Seebeck coefficient, yielding
$\tilde{k}\approx10^{-2}$\AA$^{-1}$. Comparing $\tilde{k}$ to the size of the Brillouin zone of perfect FeSb$_2$, we deduce an effective defect concentration of $9 \cdot 10^{-5}$/unit-cell.
Next, we set $\tau_i=\tau_{CB}=\tau_{VB}=0.3$ps. This fully determines the in-gap bandwidths for which we find $W_1=3$meV, $W_2=4.1$ and $W_3=5.1$ meV
\footnote{We note that a change in $\tau$ by one order of magnitude yields only a factor of 3 on the bandwidths $W_i$. Since the latter are constrained by the size of the fundamental gap $\Delta\sim30$meV,
we belief our choice to be reasonable.
}. 

With all parameters thus fixed, we now check if the value of $\alpha$ is realistic. To this end we compare the linear momentum that is transferred from the phonons to the electrons with that dissipated by lattice thermal conduction. 
We find the momentum transfer to the electrons --that is needed to justify the large $S$-- to be only $\sim 0.5\%$ of that dissipated by phonons (for details see Supplemental Material\cite{Suppl}), which is very realistic.
Indeed the experimental thermal conductivity is characterised by phonon-crystal boundary scatterings \cite{0295-5075-80-1-17008}, supporting such a small momentum transfer. 
Consequently our results show that a large phonon-drag thermopower need not produce temperature characteristics
in the thermal conductivity. 
That a minute transfer of phonon momentum has such a big impact on charge carriers is explained by
the large ratio of masses of ions and electrons.

In the Nernst coefficient, the proposed phonon-coupling to the density of in-gap states circumvents Sondheimer cancellation\cite{PhysRevB.64.224519}, without requiring multiband effects.  Previously, it had been proposed that a strongly energy-dependent scattering rate $\tau(\epsilon)$ causes the colossal Nernst signal\cite{PhysRevB.88.245203}. Indeed we can assume an {\it effective}, purely electronic picture with a scattering rate 
 $\tau(\epsilon)\propto\epsilon^r$, where the exponent is determined by $r=-e/k_B\times N/(R_H\sigma)$. This yields, in our phonon-drag formalism, a large exponent $r=-(\hbar/k_B)\alpha=-229.2$, 
in reasonable agreement with Ref.~\onlinecite{PhysRevB.88.245203}.

\begin{figure}[!t]
 \includegraphics[width=0.49\textwidth]{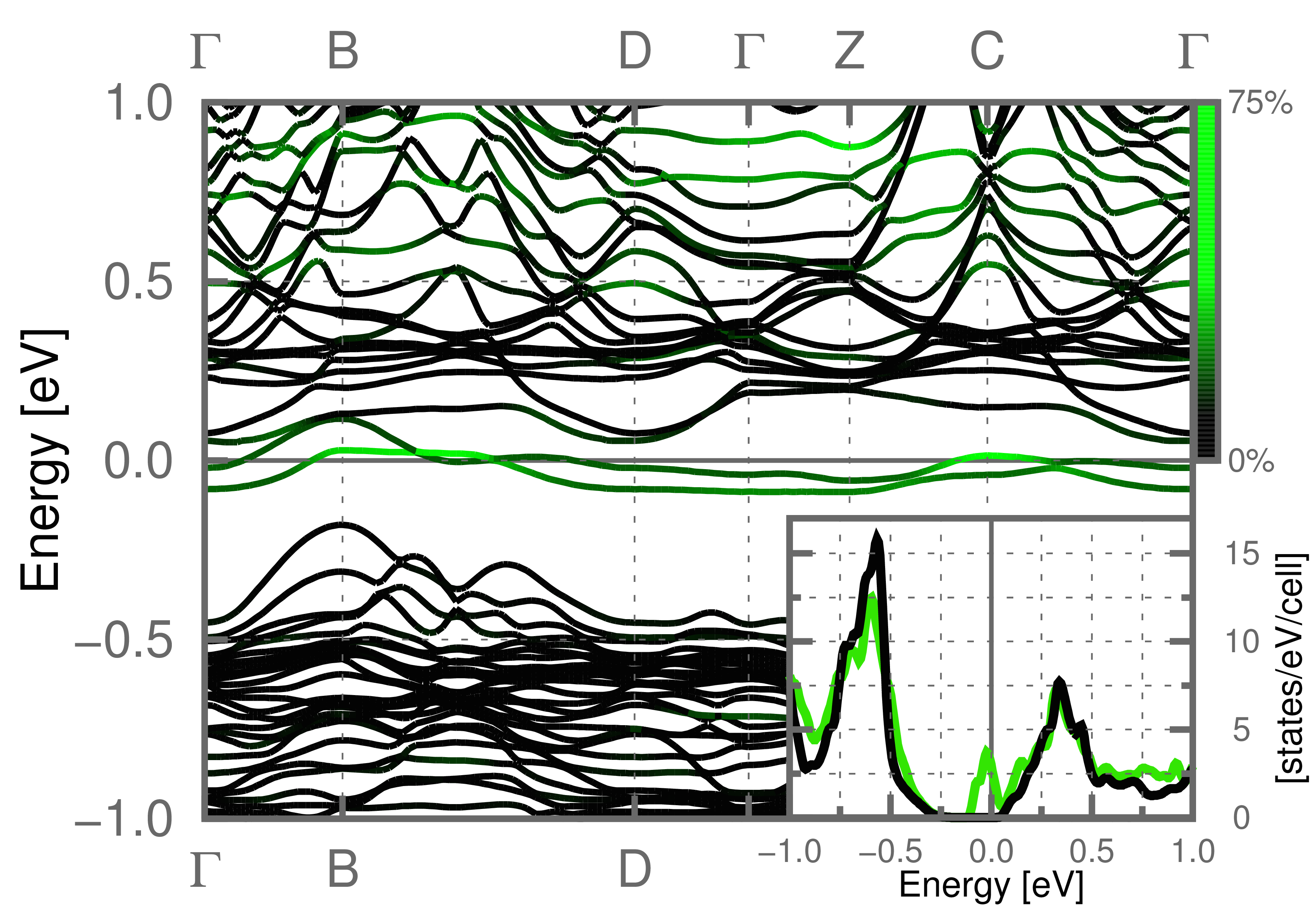}
 \caption{[Colour online] 
Band-structure of Fe$_{25}$Sb$_{47}$, mimicking Fe$_{1+x}$Sb$_{2-x}$.
Hues of green indicate orbital admixtures originating from the extra iron atom.
Inset: density of states of pure (black) and Fe-rich (green) FeSb$_2$.
}
 \label{fig:band}
\end{figure}

\paragraph{Origin of in-gap states.}
We now investigate the possible origin of the in-gap states. The number of states has to be low compared to the weight of the bottom of the conduction band, or the chemical potential will not be able to cross them. Moreover the in-gap bandwidth must be small compared to the intrinsic charge gap. This suggests the states to stem from local defects. The strong reaction to magnetic fields (MR) then points towards Fe impurities.  Using density functional theory  calculations%
, we have modelled the behavior of FeSb$_2$ with different defects and impurities. We find that substituting Fe on Sb-sites, i.e.\ Fe$_{1+x}$Sb$_{2-x}$, agrees best with the picture proposed above.  Our estimate for the minimal defect concentration (see above) translates to $x\approx 4.5\cdot 10^{-5}$\cite{Suppl}.  Simulating the effect of iron defects, we show in \fref{fig:band} the paramagnetic band structure of Fe$_{25}$Sb$_{47}$ (i.e.\ $x\approx 0.04$)\footnote{Describing the gap in FeSb$_2$ requires exchange effects beyond standard band-theory\cite{jmt_fesb2}. Here we use the modified Becke-Johnson exchange potential to mimic this physics, see Supplemental Material. Also note that a ferromagnetic solution exists, which however lies higher in energy.}: Indeed two impurity bands appear close to the conduction states. Moreover, the doping coming from the defect pins the chemical potential (at $T=0$) to the bottom of the conduction band, in extremely good agreement with the scenario proposed above.  This finding is supported by the binary Fe-Sb phase diagram\cite{Richter1997247} that may suggest Fe-richness, and a recent experiment demonstrating that even the smallest iron concentrations in Ru$_{1-x}$Fe$_x$Sb$_2$ causes the appearance of the low temperature characteristics of FeSb$_2$\cite{cava2013fesb2}.  Further support comes from recent magnetic resonance experiments\cite{Gippius_AMR} that advocate localized $S=1/2$ in-gap states  slightly below the conduction bands.   The presence of Fe defects with localized states entails a Curie-like contribution to the magnetic susceptibility, which can be estimated as follows: Curie's law reads $\chi=x \mu_0 N_A \mu_{eff}^2/(12\pi T)$, with  the vacuum permeability $\mu_0$, Avogadro's constant $N_A$, an effective local moment $\mu_{eff}=g_S \sqrt{S(S+1)}\mu_B$, with  gyromagnetic factor $g_S=2$ and Bohr magneton $\mu_B$. Using the experimentally suggested $S=1/2$ spin-state\cite{Gippius_AMR} and the minimal defect concentration $x$ yields $\chi\cdot T=5\cdot 10^{-6}$emuK/mol. 
This is consistent with experiment, witnessing  a low temperature upturn in the susceptibility for most samples, with a magnitude smaller than $5\cdot 10^{-5}$emu/mol\cite{PhysRevB.67.155205,PhysRevB.72.045103,PhysRevB.74.195130,koyama:073203,APEX.2.091102,sun_dalton,PhysRevB.83.184414} for all accessed temperatures.

\paragraph{Summary.}

Purely electronic transport properties of FeSb$_2$, like the conductivity and the Hall coefficient, can be described with impressive accuracy  from an effective one-particle electronic structure%
\footnote{
This suggests that for $T\lesssim 100$K electronic correlations renormalize but do not invalidate the band-picture. Yet, for $T\gtrsim 100$K
FeSb$_2$ metallizes\cite{perucci_optics} although $T\ll \Delta$.
A similar effect has recently been explained for FeSi\cite{jmt_fesi,jmt_hvar} (see also Ref.~\cite{Delaire22032011}), advocating strong many-body effects for $T\gtrsim100$K also in FeSb$_2$.
}
when taking into account the influence of defect-derived in-gap states. Using {\it ab initio} band-theory, we identified surplus iron, Fe$_{1+x}$Sb$_{2-x}$, as the likely origin of these states. Our work provides a unified picture: there is only one independent transport property, from which the features of all five other response functions are determined in a one-to-one correspondence.  Finally, we have pin-pointed the phonon-drag associated with the density of in-gap states as the microscopic origin of the hitherto elusive colossal thermoelectricity in FeSb$_2$. Our findings have important consequences for the quest of reducing the thermal conduction. Selective phonon engineering, e.g.\ by nano-structuring\cite{Dresselhaus2013,RevModPhys.83.131},  needs to take care that the phonon drag mechanism  survives.
Our scenario might also be relevant for understanding other narrow-gap semiconductors, such as CrSb$_2$\cite{PhysRevB.86.235136} or FeGa$_3$\cite{monika_fega3,PhysRevB.90.195206}.


\begin{acknowledgments}
We thank S.~B{\"u}hler-Paschen, Y.~Nomura, C.~Petrovic, A.~Prokofiev, F.~Steglich, P.~Sun, and V.~Zlatic for fruitful discussions,
and P.~Sun for providing the experimental data of Ref.~\cite{PhysRevB.88.245203}. This work has been supported in part by the European Research Council under the European Union's Seventh Framework Program
(FP/2007-2013)/ERC through grant agreement n.~306447,
and COST Action MP1306 EUSpec (JMT). Numerical calculations have been achieved in part using the Vienna Scientific Cluster (VSC).
\end{acknowledgments}


%

\end{document}